\documentclass[preprint,prd,aps,showpacs,showkeys,nofootinbib]{revtex4} 
\usepackage{graphicx}
\usepackage{dcolumn}
\usepackage{bm}
\begin{document}

\preprint{hep-ph/0405192}

\title{Two-loop gluino corrections to the inclusive $B\rightarrow X_{_s}\gamma$
decay in $CP$ violating MSSM with large $\tan\beta$}
 
\author{Tai-Fu Feng}

\affiliation{Department of Physics, 40014 University of Jyv\"askyl\"a,Finland}

\date{\today}

\begin{abstract}
We investigate two-loop gluino corrections to the effective Lagrangian for
$b\rightarrow s+\gamma(g)$ in the minimal supersymmetric
extension of the standard model (MSSM) at large $\tan\beta$,
including the contributions in which quark flavor change is mediated by charginos.
Using the translation invariant of loop momenta and the Ward-Takahashi 
identities (WTIs) that are required by the $SU(3)_c\times U(1)_{_{em}}$ 
gauge invariance, we simplify our expressions to concise forms. As an example,
we discuss two-loop gluino corrections to the $CP$ asymmetry of 
inclusive $B\rightarrow X_{_s}\gamma$ decay in $CP$ violating MSSM.
\end{abstract}



\maketitle

\section{Introduction}
\indent\indent
The measurements of the branching ratios at CLEO, ALEPH and BELLE
\cite{exp} give the combined result
\begin{equation}
BR(B\rightarrow X_{_s}\gamma)=(3.11\pm0.42\pm0.21)\times10^{-4}\;,
\label{eq1}
\end{equation}
which agrees with the next-to-leading order (NLO) standard model
(SM) prediction \cite{smp}
\begin{equation}
BR(B\rightarrow X_{_s}\gamma)=(3.29\pm0.33)\times10^{-4}\;.
\label{eq2}
\end{equation}
Good agreement between the experiment and the theoretical prediction
of the SM implies that the new physics scale should lie well above
the electroweak (EW) scale. The systematic analysis of
new physics corrections to $B\rightarrow X_{_s}\gamma$
up to two-loop order can help us understanding where the
new physics scale sets in, and the distribution of new physical
particle masses around this scale. In principle, the two-loop
corrections can be large when some additional parameters are involved
at this perturbation order beside the parameters appearing in
one loop results. In other words, including the two-loop
contributions one can obtain a more exact constraint on the new physics
parameter space from the present experimental results.

Beside the Cabibbo-Kobayashi-Maskawa (CKM) mechanism, the soft breaking
terms provide a new source of $CP$ and flavor violation
in the MSSM. Those $CP$ violating phases can affect the important
observables in the mixing of Higgs bosons \cite{higgs},
the lepton and neutron's electric dipole moments (EDMs) \cite{edm1, edm2},
lepton polarization asymmetries in the semi-leptonic decays \cite{semi-lep},
the production of $P$-wave charmonium and bottomonium \cite{quarknium},
and $CP$ violation in rare $B$-decays and in $B^0\bar{B}^0$
mixing \cite{dmix}. At present, the strictest constraints on those $CP$ violation
phases originate from the lepton and neutron's EDMs. Nevertheless,
if we invoke a cancellation mechanism among different supersymmetric
contributions \cite{edm1}, or choose the sfermions of the first generation
heavy enough \cite{edm2}, the loop inducing lepton and neutron's EDMs
bound the argument of the $\mu$ parameter to be $\le \pi/(5\tan\beta)$,
leaving no constraints on the other explicitly $CP$ violating phases.

The supersymmetry models at large $\tan\beta$ are implied
by grand unified theories, where the unification of up- and down-type
quark Yukawa couplings is made \cite{uni}. From the technical viewpoint,
the dominant contributions to the relevant effective Lagrangian are
the terms proportional to $(\tan\beta)^n\;(n=1,\;2,\cdots)$ 
in a large $\tan\beta$ scenario.
This will simplify our two-loop analysis drastically since we just keep those
terms enhanced by $\tan\beta$.

Assuming no additional sources of flavor violation other than
the CKM matrix elements, the authors of \cite{Borzumati}
present an exact analysis of the two-loop gluino corrections to
the rare decay $b\rightarrow s+\gamma(g)$ in which quark
flavor change is mediated by the charged Higgs in $CP$
conserving MSSM at large $\tan\beta$.
They also  compare their exact result with that originating from
the heavy mass expansion (HME) approximation \cite{hme}. Although
the HME result approximates the exact two-loop analysis adequately when
the supersymmetry energy scale is high enough, their analysis
implies that the difference between the HME approximation and exact
calculation is obvious in some parameter space of the MSSM.
However, they do not consider the case in which quark flavor change is
mediated by the charginos (the super partners of the charged Higgs 
and $W$ bosons). In fact, we cannot provide
any strong reason to ignore the contribution from the diagrams
in which quark flavor change is induced by the charginos, even within
large $\tan\beta$ scenarios. In this work, we present a complete 
analysis on the two-loop gluino corrections to the rare transitions 
$b\rightarrow s+\gamma(g)$ by including the contributions of those 
diagrams where quark flavor change is mediated by charginos in the
framework of $CP$ violating MSSM at large $\tan\beta$. 
Furthermore, we also simplify our expressions
to concise forms through loop momentum invariant
and the WTIs that are required by the $SU(3)_c\times U(1)_{em}$ 
gauge invariance.

The paper is organized as follows. In Sec. \ref{dia}, we give
all the diagrams needed to evaluate the $O(\alpha_s\tan\beta)$
contributions to the Wilson coefficients $C_{_7}$ and $C_{_8}$
entering the branching ratio $BR(B\rightarrow X_{_s}\gamma)$.
The corresponding Wilson coefficients at the matching EW scale
$\mu_{_{\rm EW}}$ are also presented there. We apply the effective
Lagrangian to the rare decay $B\rightarrow X_{_s}\gamma$ in Sec.
\ref{cp}. By the numerical method, we show the two-loop corrections
on the $CP$ asymmetry for the process. Our conclusion is given 
in Sec. \ref{con}, and some long formulae are collected in appendices.

\section{The Wilson coefficients from the two-loop diagrams\label{dia}}
\indent\indent

In this section, we derive the relevant Wilson coefficients
for the partonic decay $b\rightarrow s\gamma$ including two-loop
gluino corrections. In a conventional form, the effective Hamilton is
written as
\begin{eqnarray}
&&H_{_{eff}}= -{4G_{_F}\over\sqrt{2}}V_{_{ts}}^*V_{_{tb}}
\sum\limits_{i=1}^8C_{_i}(\mu){\cal O}_{_i}\;,
\label{eq3}
\end{eqnarray}
where $V$ is the CKM matrix. The definitions of the magnetic and
chromo-magnetic dipole operators are
\begin{eqnarray}
&&{\cal O}_{_7}={e\over(4\pi)^2}m_{_b}(\mu)\bar{s}_{_L}\sigma^{\mu\nu}
b_{_R}F_{_{\mu\nu}}\;,\nonumber\\
&&{\cal O}_{_8}={g_{_s}\over(4\pi)^2}m_{_b}(\mu)\bar{s}_{_L}T^a\sigma^{\mu\nu}
b_{_R}G^a_{_{\mu\nu}}\;,
\label{eq4}
\end{eqnarray}
where $F_{_{\mu\nu}}$ and $G^a_{_{\mu\nu}}$ are the field strengths of the
photon and gluon respectively, and $T^a\;(a=1,\;\cdots,\;8)$ are $SU(3)_{_c}$
generators. In addition, $e$ and $g_{_s}$ represent the EW and strong couplings
respectively. The other operators ${\cal O}_{_i}\;(i=1,\;\cdots,\;6)$
are defined in \cite{eh}. 

\begin{figure}[t]
\setlength{\unitlength}{1mm}
\begin{center}
\begin{picture}(0,20)(0,0)
\put(-90,-230){\includegraphics{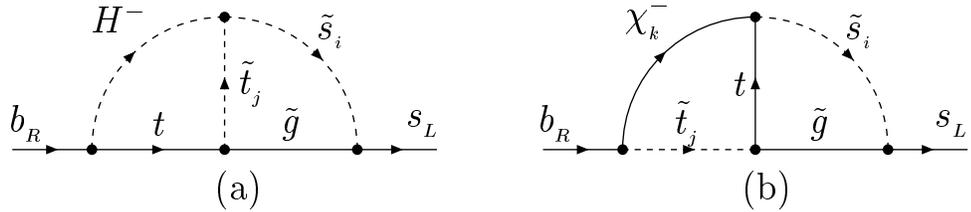}}
\end{picture}
\caption[]{The self energy diagrams which lead to the magnetic and
chromo-magnetic operators in the MSSM, the corresponding triangle
diagrams are obtained by attaching a photon or  gluon in all possible
ways.} \label{fig1}
\end{center}
\end{figure}

In the framework of $CP$ violation MSSM, the one-loop analysis on the 
$CP$ asymmetry of inclusive $B\rightarrow X_{_s}\gamma$ decay has been 
presented elsewhere \cite{uni}. The two loop gluino diagrams, 
contributing at ${\cal O}(\alpha_s\tan\beta)$
to the Wilson coefficients of the magnetic and chromo-magnetic dipole
operators, are obtained from the self energy diagrams $(a),\;(b)$
of FIG. \ref{fig1} by attaching a photon or gluon in all possible
ways. The calculation of
the Wilson coefficients for the operators in Eq. (\ref{eq4})
at the two loop order is more challenging than that
at the one loop order. 
Before we give those Wilson coefficient expressions
explicitly, we state firstly the concrete steps required to obtain the coefficients
from those two loop diagrams. 
\begin{itemize}
\item After writing the amplitudes of those two loop triangle diagrams,
we expand them in powers of external momenta to the second
order. 
\begin{figure}
\setlength{\unitlength}{1mm}
\begin{center}
\begin{picture}(0,50)(0,0)
\put(-80,-135){\includegraphics{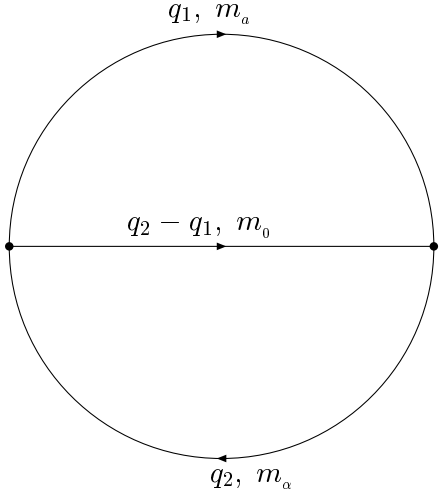}}
\end{picture}
\caption[]{The two-loop vacuum diagram with momenta and masses as 
in Eq. \ref{eq6}.} \label{vac}
\end{center}
\end{figure}
\item The even rank tensors in the loop momenta $q_1,\;q_2$ can be replaced
as follows
\begin{widetext}
\begin{eqnarray}
&&\int{d^Dq_1\over(2\pi)^D}{d^Dq_2\over(2\pi)^D}{q_{1\mu}q_{1\nu},\;
q_{1\mu}q_{2\nu}\over{\cal D}_{_0}}
\longrightarrow {g_{_{\mu\nu}}\over D}\int{d^Dq_1\over(2\pi)^D}
{d^Dq_2\over(2\pi)^D}{q_1^2,\;q_1\cdot q_2\over{\cal D}_{_0}}
\;,\nonumber\\
&&\int{d^Dq_1\over(2\pi)^D}{d^Dq_2\over(2\pi)^D}{q_{1\mu}q_{1\nu}q_{1\rho}
q_{1\sigma},\;q_{1\mu}q_{1\nu}q_{1\rho}q_{2\sigma}\over{\cal D}_{_0}}
\nonumber\\&&
\longrightarrow {T_{_{\mu\nu\rho\sigma}}\over D(D+2)}\int{d^Dq_1\over(2\pi)^D}
{d^Dq_2\over(2\pi)^D}{q_1^4,\;q_1^2(q_1\cdot q_2)
\over{\cal D}_{_0}}\;,\nonumber\\
&&\int{d^Dq_1\over(2\pi)^D}{d^Dq_2\over(2\pi)^D}{q_{1\mu}q_{1\nu}q_{2\rho}
q_{2\sigma}\over{\cal D}_{_0}}
\nonumber\\&&
\longrightarrow\int{d^Dq_1\over(2\pi)^D}
{d^Dq_2\over(2\pi)^D}{1\over{\cal D}_{_0}}\Big({D(q_1\cdot q_2)^2-q_1^2q_2^2\over D(D-1)(D+2)}
T_{_{\mu\nu\rho\sigma}}
-{(q_1\cdot q_2)^2-q_1^2q_2^2\over D(D-1)}
g_{_{\mu\nu}}g_{_{\rho\sigma}}\Big)\;,
\label{eq6}
\end{eqnarray}
\end{widetext}
where $D$ is the time-space dimension, 
$T_{_{\mu\nu\rho\sigma}}=g_{_{\mu\nu}}g_{_{\rho\sigma}}
+g_{_{\mu\rho}}g_{_{\nu\sigma}}+g_{_{\mu\sigma}}g_{_{\rho\nu}}$
and ${\cal D}_{_0}=((q_2-q_1)^2-m_{_0}^2)(q_1^2-m_{_a}^2)
(q_2^2-m_{_\alpha}^2)$.
The odd rank tensors in the loop momenta can be dropped since
the integrations are symmetric under the transformation $q_{1,2}
\rightarrow -q_{1,2}$. Here, we only retain
the simplest two-loop propagator composition $1/{\cal D}_{_0}$
which corresponds to the two-loop vacuum diagram (FIG. \ref{vac}). 
Any complicated composition of two-loop propagators 
can be expressed as the linear combination of the simplest one $1/{\cal D}_{_0}$
by use of the obvious decomposition formula
\begin{eqnarray}
&&{1\over(Q^2-m_{_A}^2)(Q^2-m_{_B}^2)}={1\over m_{_A}^2-m_{_B}^2}
\Big({1\over Q^2-m_{_A}^2}-{1\over Q^2-m_{_B}^2}\Big)\;,
\label{exp2}
\end{eqnarray}
with $Q=q_1,\;q_2$, or $q_2-q_1$.
\begin{figure}
\setlength{\unitlength}{1mm}
\begin{center}
\begin{picture}(0,50)(0,0)
\put(-60,-135){\includegraphics{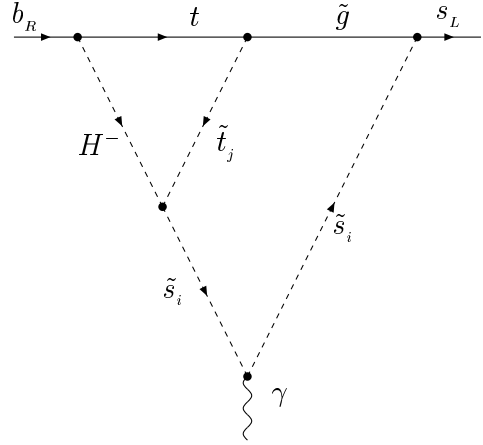}}
\end{picture}
\caption[]{A triangle diagram in which the external photon
is attached to squark $\tilde{s}_{_i}$.} \label{example}
\end{center}
\end{figure}
As an example, we apply the above two steps to the triangle diagram
in which a external photon is attached to the internal 
squark $\tilde{s}_{_i}$ line (FIG. \ref{example}). After expanding
the corresponding amplitude in powers of external momenta to the
second order, we have
\begin{widetext}
\begin{eqnarray}
&&iA^{\gamma}_{\mu}(p,k)
=-i{4\over3}e_{_d}g_s^2{e^3\over m_{_{\rm w}}s_{_{\rm w}}^2}V_{_{ts}}^*V_{_{tb}}
({m_{_b}\over m_{_{\rm w}}}\tan\beta)({\cal Z}_{_{\tilde s}})_{_{2,i}}
({\cal Z}_{_{\tilde s}}^\dagger)_{_{i,2}}\big(|\mu|m_{_{t}}e^{-i\theta_{_\mu}}
({\cal Z}_{_{\tilde t}})_{_{2,j}}
\nonumber\\
&&\hspace{2.2cm}
+{\sqrt{2}m_{_{\rm w}}s_{_{\rm w}}{\bf A}_{_s}\over e}
({\cal Z}_{_{\tilde t}})_{_{1,j}}\big)
\int{d^Dq_1\over(2\pi)^D}{d^Dq_2\over(2\pi)^D}
{1\over{\cal D}_{_H}(q_1^2-m_{_{\tilde{s}_i}}^2)}
\nonumber\\
&&\hspace{2.2cm}\times
\Big\{1+{2q_1\cdot(2p+k)\over q_1^2-m_{_{\tilde{s}_i}}^2}
+{2q_2\cdot p\over q_2^2-m_{_{H^+}}^2}\Big\}
\nonumber\\
&&\hspace{2.2cm}\times
\Big\{({\cal Z}_{_{\tilde t}}^\dagger)_{_{j,3}}
\;/\!\!\!q_1/\!\!\!q_2(2q_1-2p-k)_\mu\omega_+
\nonumber\\
&&\hspace{2.2cm}
-m_{_t}|m_{_3}|e^{i\theta_{_3}}({\cal Z}_{_{\tilde t}}^\dagger)_{_{j,2}}
\;(2q_1-2p-k)_\mu\omega_+\Big\}\;,
\label{exp1}
\end{eqnarray}
\end{widetext}
where $p,\;k$ represent the incoming momenta of the external quark $b$
and photon respectively, ${\cal D}_{_H}=((q_2-q_1)^2-m_{_{\tilde{t}_j}}^2)(q_1^2-|m_{_3}|^2)
(q_1^2-m_{_{\tilde{s}_i}}^2)(q_2^2-m_{_t}^2)(q_2^2-m_{_{H^+}}^2)\;(i,\;j=1,\;2)$ and $e_{_d}=-1/3$. 
${\cal Z}_{_{\tilde q}}\;(q=u,\;d,\;\cdots,\;t)$ are the mixing matrices
of scalar quarks, and ${\bf A}_{_{q}}$ are the corresponding trilinear
soft breaking parameters. Furthermore, $\theta_{_{3,\mu}}$ denote the $CP$ phases
of the $SU(3)_{_c}$ gaugino mass and of the $\mu$ parameter respectively.
Since the quark mass $m_{_b}$ from the Yukawa coupling of bottom quark
is same order as the external momenta $p,\;k$ in magnitude, we just
expand the propagators in powers of the external momenta to the first order.
In the soft breaking potential, the $CP$ phase $\theta_{_{3}}$ is contained 
in the gluino mass terms
\begin{eqnarray}
&&|m_{_3}|e^{i\theta_{_3}}\lambda_{_G}\lambda_{_G}+|m_{_3}|e^{-i\theta_{_3}}
\overline{\lambda}_{_G}\overline{\lambda}_{_G}\;,
\label{exp1a}
\end{eqnarray}
where $\lambda_{_G}$ denotes the gluino in a two-component Majorana spinor. With
the redefinition of the gluino field 
\begin{eqnarray}
&&\lambda_{_G}\rightarrow \lambda_{_G}e^{-{i\over2}\theta_{_3}}\;,
\nonumber\\
&&\overline{\lambda}_{_G}\rightarrow \overline{\lambda}_{_G}
e^{{i\over2}\theta_{_3}}\;,
\label{exp1b}
\end{eqnarray}
the mass terms are transformed into
\begin{eqnarray}
&&|m_{_3}|\overline{\tilde{g}}\tilde{g}
\label{exp1c}
\end{eqnarray}
with the four-component Majorana spinor
\begin{eqnarray}
&&\tilde{g}=\left(\begin{array}{c}\lambda_{_G}\\\overline{\lambda}_{_G}
\end{array}\right)\;.
\label{exp1d}
\end{eqnarray}
Correspondingly, the $CP$ phase $\theta_{_{3}}$ is transfered from the
mass terms to the quark-squark-gluino vertex which is given by \cite{add1}
\begin{eqnarray}
&&-{\cal L}_{_{\tilde{q}q\tilde{g}}}=\sqrt{2}g_{_s}T^a_{\alpha\beta}
\sum\limits_{q}\Big[-e^{-{i\over2}\theta_{_3}}({\cal Z}_{_{\tilde q}})_{_{2,i}}
\overline{q}^\alpha\omega_-\tilde{g}_a\tilde{q}_{_i}^\beta
\nonumber\\
&&\hspace{1.6cm}+e^{{i\over2}\theta_{_3}}({\cal Z}_{_{\tilde q}})_{_{1,i}}
\overline{q}^\alpha\omega_+\tilde{g}_a\tilde{q}_{_i}^\beta\Big]+{\rm H.c.}
\label{exp1e}
\end{eqnarray}
Here, $\alpha,\;\beta=1,\;2,\;3$ are quark and squark color indices, and $\omega_\pm=
{1\pm\gamma_5\over2}$. This is the reason why there is a $|m_{_3}|$
rather than $m_{_3}$ in the gluino propagator.
Using Eq. \ref{eq6} and Eq. \ref{exp2}, the amplitude of FIG. \ref{example} 
is finally formulated as
\begin{eqnarray}
&&iA^{\gamma}_{\mu}(p,k)
=-i{4\over3}e_{_d}g_s^2{e^3\over m_{_{\rm w}}s_{_{\rm w}}^2}V_{_{ts}}^*V_{_{tb}}
({m_{_b}\over m_{_{\rm w}}}\tan\beta)({\cal Z}_{_{\tilde s}})_{_{2,i}}
({\cal Z}_{_{\tilde s}}^\dagger)_{_{i,2}}\big(|\mu|m_{_{t}}e^{-i\theta_{_\mu}}
({\cal Z}_{_{\tilde t}})_{_{2,j}}
\nonumber\\
&&\hspace{2.2cm}
+{\sqrt{2}m_{_{\rm w}}s_{_{\rm w}}{\bf A}_{_s}\over e}
({\cal Z}_{_{\tilde t}})_{_{1,j}}\big)
\int{d^Dq_1\over(2\pi)^D}{d^Dq_2\over(2\pi)^D}
{1\over{\cal D}_{_H}(q_1^2-m_{_{\tilde{s}_i}}^2)}
\nonumber\\
&&\hspace{2.2cm}\times
\Big\{({\cal Z}_{_{\tilde t}}^\dagger)_{_{j,3}}
\Big[{4\over D}{q_1^2q_1\cdot q_2
\over q_1^2-m_{_{\tilde{s}_i}}^2}\;(2p+k)_\mu\omega_+
-q_1\cdot q_2\;(2p+k)_\mu\omega_+
\nonumber\\
&&\hspace{2.2cm}
+{4\over q_2^2-m_{_{H^+}}^2}\Big({D(q_1\cdot q_2)^2-q_1^2q_2^2\over D(D-1)}
\;p_\mu\omega_+
-{(q_1\cdot q_2)^2-q_1^2q_2^2\over D(D-1)}\;\gamma_\mu/\!\!\!p
\omega_+\Big)\Big]
\nonumber\\
&&\hspace{2.2cm}
-m_{_t}|m_{_3}|e^{i\theta_{_3}}
({\cal Z}_{_{\tilde t}}^\dagger)_{_{j,2}}
\Big[{4\over D}{q_1^2\over q_1^2-m_{_{\tilde{s}_i}}^2}\;(2p+k)_\mu\omega_+
+{4\over D}{q_1\cdot q_2\over q_2^2-m_{_{H^+}}^2}\;p_\mu\omega_+
\nonumber\\
&&\hspace{2.2cm}
-\;(2p+k)_\mu\omega_+\Big]\Big\}\;.
\label{exp3}
\end{eqnarray}
In a similar way, we can obtain the other triangle diagram amplitudes.
\item Using loop momentum translation invariant, we formulate
the sum of those amplitudes in gauge invariance form explicitly,
then extract the corresponding Wilson coefficients which are
expressed by the two-loop vacuum integrals \cite{Davydychev}.
In fact, there are many identities among those two-loop integrations.
In order to obtain those necessary identities which are used to
simplify the sum of those triangle amplitudes, we start
from the zero integration such as
\begin{eqnarray}
&&\int{d^Dq_1\over(2\pi)^D}{d^Dq_2\over(2\pi)^D}{q_1\cdot p
\;/\!\!\!q_1(\;/\!\!\!q_2-\;/\!\!\!q_1)\over{\cal D}_{_0}}
\equiv0\;.
\label{exp4}
\end{eqnarray}
Under the loop momentum translation $q_2\rightarrow q_2-a$ where $a$
is same order as the external momentum $p$ in magnitude, we expand the integration in powers of
the momenta $p,\;a$ to the second order
\begin{eqnarray}
&&\int{d^Dq_1\over(2\pi)^D}{d^Dq_2\over(2\pi)^D}{q_1\cdot p\over{\cal D}_{_0}}
\;/\!\!\!q_1(\;/\!\!\!q_2-\;/\!\!\!q_1)
\nonumber\\&&
=\int{d^Dq_1\over(2\pi)^D}{d^Dq_2\over(2\pi)^D}{q_1\cdot p\over{\cal D}_{_0}
}\Big\{1+{2(q_2-q_1)\cdot a\over (q_2-q_1)^2-m_{_0}^2}
+{2q_2\cdot a\over q_2^2-m_{_\alpha}^2}\Big\}
\nonumber\\&&\times
\Big\{/\!\!\!q_1(\;/\!\!\!q_2-\;/\!\!\!q_1)-/\!\!\!q_1/\!\!\!a\Big\}
\nonumber\\&&
=\int{d^Dq_1\over(2\pi)^D}{d^Dq_2\over(2\pi)^D}{1\over{\cal D}_{_0}}
\Big\{-{q_1^2\over D}/\!\!\!p/\!\!\!a+{2\over (q_2-q_1)^2-m_{_0}^2}
\Big[\big({D(q_1\cdot q_2)^2-q_1^2q_2^2\over D(D-1)}
\nonumber\\&&
+{q_1^4-2q_1^2q_1\cdot q_2\over D}\big)(p\cdot a)
-{(q_1\cdot q_2)^2-q_1^2q_2^2\over D(D-1)}(/\!\!\!p/\!\!\!a)\Big]
\nonumber\\&&
+{2\over q_2^2-m_{_\alpha}^2}\Big[\big({D(q_1\cdot q_2)^2-q_1^2q_2^2\over D(D-1)}
-{q_1^2q_1\cdot q_2\over D}\big)(p\cdot a)
\nonumber\\&&
-{(q_1\cdot q_2)^2-q_1^2q_2^2\over D(D-1)}
(/\!\!\!p/\!\!\!a)\Big]\Big\}\equiv0\;.
\label{exp5}
\end{eqnarray}
The above identical equation implies
\begin{eqnarray}
&&\int{d^Dq_1\over(2\pi)^D}{d^Dq_2\over(2\pi)^D}{1\over{\cal D}_{_0}}
\Big\{{1\over (q_2-q_1)^2-m_{_0}^2}
\big[{D(q_1\cdot q_2)^2-q_1^2q_2^2\over D(D-1)}+{q_1^4-2q_1^2q_1\cdot q_2\over D}\big]
\nonumber\\&&
+{1\over q_2^2-m_{_\alpha}^2}\big[{D(q_1\cdot q_2)^2-q_1^2q_2^2\over D(D-1)}
-{q_1^2q_1\cdot q_2\over D}\big]\Big\}\equiv0\;,
\nonumber\\&&
\int{d^Dq_1\over(2\pi)^D}{d^Dq_2\over(2\pi)^D}{1\over{\cal D}_{_0}}
\Big\{{2\over (q_2-q_1)^2-m_{_0}^2}{(q_1\cdot q_2)^2-q_1^2q_2^2
\over D(D-1)}
\nonumber\\&&
+{2\over q_2^2-m_{_\alpha}^2}{(q_1\cdot q_2)^2-q_1^2q_2^2
\over D(D-1)}+{q_1^2\over D}\Big\}\equiv0\;.
\label{exp6}
\end{eqnarray}
Similarly, we can get the following identities from the invariant of Eq. \ref{exp4}
under the loop momentum translation $q_1\rightarrow q_1-a,\;q_2\rightarrow q_2-a$:
\begin{eqnarray}
&&\int{d^Dq_1\over(2\pi)^D}{d^Dq_2\over(2\pi)^D}{1\over{\cal D}_{_0}}
\Big\{-{2+D\over D}q_1\cdot(q_2-q_1)
+{2\over q_1^2-m_{_a}^2}{q_1^2q_1\cdot(q_2-q_1)\over D}
\nonumber\\&&
+{2\over q_2^2-m_{_\alpha}^2}\Big[{D(q_1\cdot q_2)^2-q_1^2q_2^2\over D(D-1)}
-{q_1^2q_1\cdot q_2\over D}\Big]\Big\}\equiv0\;,
\nonumber\\&&
\int{d^Dq_1\over(2\pi)^D}{d^Dq_2\over(2\pi)^D}{1\over{\cal D}_{_0}}
\Big\{{q_1\cdot(q_2-q_1)\over D}
-{2\over q_2^2-m_{_\alpha}^2}{(q_1\cdot q_2)^2-q_1^2q_2^2\over D(D-1)}\Big\}\equiv0\;.
\label{exp7}
\end{eqnarray}
Using the concrete expressions of two-loop vacuum integrals in Ref. \cite{Davydychev},
we can also verify those equations in Eq. \ref{exp6} and Eq. \ref{exp7} directly
after some tedious calculations. Replacing the numerator of Eq. \ref{exp4} with other
odd rank tensors in the loop momenta $q_1,\;q_2$, the more additional identities
among two-loop integrations are gotten. In general, those identities are linearly dependent.
After some simplification, we obtain those linearly independent equations in appendix \ref{appa}.
Certainly, those linearly independent equations can also be derived from those two-loop
integrations in which the numerators are even rank tensors of the loop momenta
$q_1,\;q_2$. However, the process to derive the linearly independent equations 
with the numerators in even powers of the loop momenta is more complicated than 
that with the numerators in odd powers of the loop momenta.
\end{itemize}

After the above procedure, we finally obtain the relevant coefficients
from the charged Higgs contribution up to ${\cal O}(\alpha_{_s}\tan\beta)$
\begin{widetext}
\begin{eqnarray}
&&C_{_{7,H}}(\mu_{_{\rm w}})={8\sqrt{2}\over3}(4\pi)^3e_{_d}
(\alpha_{_s}\tan\beta)({\cal Z}_{_{\tilde s}})_{_{2,i}}
({\cal Z}_{_{\tilde s}}^\dagger)_{_{i,2}}\Big(
|\mu|m_{_t}e^{-i\theta_{_\mu}}
({\cal Z}_{_{\tilde t}})_{_{2,j}}
\nonumber\\
&&\hspace{2.2cm}
+{\sqrt{2}s_{_{\rm w}}m_{_{\rm w}}
A_{_s}\over e}({\cal Z}_{_{\tilde t}})_{_{1,j}}\Big)
\int{d^4q_1\over(2\pi)^4}{d^4q_2\over(2\pi)^4}{1\over {\cal D}_{_H}}
\Big\{({\cal Z}_{_{\tilde t}}^\dagger)_{_{j,1}}{\cal N}_{_{H(1)}}^\gamma
\nonumber\\
&&\hspace{2.2cm}
-m_{_t}|m_{_3}|e^{i\theta_{_3}}({\cal Z}_{_{\tilde t}}^\dagger)_{_{j,2}}
{\cal N}_{_{H(2)}}^\gamma\Big\}
\;,\nonumber\\
&&C_{_{8,H}}(\mu_{_{\rm w}})={8\sqrt{2}\over3}(4\pi)^3
(\alpha_{_s}\tan\beta)({\cal Z}_{_{\tilde s}})_{_{2,i}}
({\cal Z}_{_{\tilde s}}^\dagger)_{_{i,2}}\Big(
|\mu|m_{_t}e^{-i\theta_{_\mu}}
({\cal Z}_{_{\tilde t}})_{_{2,j}}
\nonumber\\
&&\hspace{2.2cm}
+{\sqrt{2}s_{_{\rm w}}m_{_{\rm w}}
A_{_s}\over e}({\cal Z}_{_{\tilde t}})_{_{1,j}}\Big)
\int{d^4q_1\over(2\pi)^4}{d^4q_2\over(2\pi)^4}{1\over {\cal D}_{_H}}
\Big\{({\cal Z}_{_{\tilde t}}^\dagger)_{_{j,1}}{\cal N}_{_{H(1)}}^g
\nonumber\\
&&\hspace{2.2cm}
-m_{_t}|m_{_3}|e^{i\theta_{_3}}({\cal Z}_{_{\tilde t}}^\dagger)_{_{j,2}}
{\cal N}_{_{H(2)}}^g\Big\}\;,
\label{eq7}
\end{eqnarray}
\end{widetext}
where $\alpha_{_s}=g_s^2/4\pi$, and the expressions of the form factors 
${\cal N}_{_{H(1,2)}}^{\gamma,\;g}$ can be found in appendix \ref{appb}.
Note, Ref. \cite{Borzumati} has also obtained the Wilson coefficients from
the same diagrams. We formulate our expressions in the more concise
forms using the identities from appendix \ref{appa}. For the chargino contribution
that is ignored by Ref. \cite{Borzumati}, we can similarly have
\begin{widetext}
\begin{eqnarray}
&&C_{_{7,\chi_k}}(\mu_{_{\rm w}})={8\sqrt{2}\over3}e_{_d}(4\pi)^3
\Big(\alpha_{_s}\tan\beta\Big)
({\cal Z}_{_{\tilde s}})_{_{2,i}}({\cal Z}_{_{\tilde s}}^\dagger)_{_{i,2}}
({\cal Z}_{_{\tilde t}}^\dagger)_{_{j,1}}({\cal Z}_-^\dagger)_{_{k,2}}
\nonumber\\
&&\hspace{2.2cm}\times
\int{d^4q_1\over(2\pi)^4}{d^4q_2\over(2\pi)^4}{1\over{\cal D}_{_{\chi_k}}}
\Big\{m_{_t}m_{_{\chi_k}}({\cal Z}_{_{\tilde t}})_{_{2,j}}({\cal Z}_+)_{_{2,k}}
{\cal N}_{_{\chi_k^\pm(1)}}^\gamma
\nonumber\\
&&\hspace{2.2cm}
+\sqrt{2}m_{_{\rm w}}m_{_t}({\cal Z}_{_{\tilde t}})_{_{2,j}}({\cal Z}_-)_{_{1,k}}
{\cal N}_{_{\chi_k^\pm(2)}}^\gamma
\nonumber\\
&&\hspace{2.2cm}
-\sqrt{2}m_{_{\rm w}}|m_{_3}|e^{i\theta_{_3}}({\cal Z}_{_{\tilde t}})_{_{1,j}}({\cal Z}_-)_{_{1,k}}
{\cal N}_{_{\chi_k^\pm(3)}}^\gamma
\nonumber\\
&&\hspace{2.2cm}
+m_{_t}^2m_{_{\chi_k}}|m_{_3}|e^{i\theta_{_3}}
({\cal Z}_{_{\tilde t}})_{_{1,j}}({\cal Z}_+)_{_{2,k}}
{\cal N}_{_{\chi_k^\pm(4)}}^\gamma\Big\}\;,\nonumber\\
&&C_{_{8,\chi_k}}(\mu_{_{\rm w}})={8\sqrt{2}\over3}(4\pi)^3
\Big(\alpha_{_s}\tan\beta\Big)
({\cal Z}_{_{\tilde s}})_{_{2,i}}({\cal Z}_{_{\tilde s}}^\dagger)_{_{i,2}}
({\cal Z}_{_{\tilde t}}^\dagger)_{_{j,1}}({\cal Z}_-^\dagger)_{_{k,2}}
\nonumber\\
&&\hspace{2.2cm}\times
\int{d^4q_1\over(2\pi)^4}{d^4q_2\over(2\pi)^4}{1\over{\cal D}_{_{\chi_k}}}
\Big\{m_{_t}m_{_{\chi_k}}
({\cal Z}_{_{\tilde t}})_{_{2,j}}({\cal Z}_+)_{_{2,k}}
{\cal N}_{_{\chi_k^\pm(1)}}^g\nonumber\\
&&\hspace{2.2cm}
+\sqrt{2}m_{_{\rm w}}m_{_t}({\cal Z}_{_{\tilde t}})_{_{2,j}}({\cal Z}_-)_{_{1,k}}
{\cal N}_{_{\chi_k^\pm(2)}}^g
\nonumber\\
&&\hspace{2.2cm}
-\sqrt{2}m_{_{\rm w}}|m_{_3}|e^{i\theta_{_3}}
({\cal Z}_{_{\tilde t}})_{_{1,j}}({\cal Z}_-)_{_{1,k}}
{\cal N}_{_{\chi_k^\pm(3)}}^g
\nonumber\\
&&\hspace{2.2cm}
+m_{_t}^2m_{_{\chi_k}}|m_{_3}|e^{i\theta_{_3}}
({\cal Z}_{_{\tilde t}})_{_{1,j}}({\cal Z}_+)_{_{2,k}}
{\cal N}_{_{\chi_k^\pm(4)}}^g\Big\}\;,
\label{eq8}
\end{eqnarray}
\end{widetext}
with ${\cal D}_{_{\chi_k}}=((q_2-q_1)^2-m_{_t}^2)(q_1^2-|m_{_3}|^2)
(q_1^2-m_{_{\tilde{s}_i}}^2)(q_2^2-m_{_{\tilde{t}_j}}^2)
(q_2^2-m_{_{\chi_k}}^2)$. ${\cal Z}_-,\;{\cal Z}_+$ are the left-
and right-handed mixing matrices of charginos, and the form factors
${\cal N}_{_{\chi_k^\pm(i)}}^{\gamma,\;g}\;(i=1,\;2,\;3,\;4)$ are 
collected in appendix \ref{appb}. After we simplify 
the sum of the $\bar{s}b\gamma\;(g)$ triangle diagram amplitudes 
using the identities in appendix \ref{appa}, we find that the effective
$\bar{s}b\gamma\;(g)$ vertices should also include the two-point
operator
\begin{eqnarray}
&&{\cal O}_{_{\rm se}}={1\over(4\pi)^2}m_{_b}(\mu)\bar{s}_{_L}(i/\!\!\!\!D)^2
b_{_R}\;,
\label{eq5}
\end{eqnarray}
beside the magnetic (chromo-magnetic) dipole operators ${\cal O}_{_{7}}\;
({\cal O}_{_{8}})$. Here, the covariant derivative acting on the quark
fields is
\begin{eqnarray}
&&D_{_\mu}=\partial_{_\mu}-iee_{_q}A_{_\mu}-ig_{_s}G_{_\mu},\;
\label{adeq1}
\end{eqnarray}
with $G_{_\mu}=G_{_\mu}^aT^a$.
Certainly, the Wilson coefficient of this operator does not give
any contribution to the rare process $b\rightarrow s\gamma$ after
we evolve the corresponding coefficients from the matching EW scale 
to the hadronic scale. Nevertheless, when we extract the Wilson coefficients
of ${\cal O}_{_{7}}\;({\cal O}_{_{8}})$, it makes sense to keep this 
operator for the following reason. 
Beside the effective vertex with two quarks
\begin{eqnarray}
&&{\cal O}_{_{\rm se}}\sim {i\over(4\pi)^2}m_{_b}p^2\omega_+\;,
\label{adeq2}
\end{eqnarray}
the operator ${\cal O}_{_{\rm se}}$ can also induce the effective vertices 
with two quarks and one photon or gluon
\begin{eqnarray}
&&{\cal O}_{_{\rm se}}\sim {i\over(4\pi)^2}ee_{_d}m_{_b}\Big(2p_{_\mu}
+/\!\!\!k\gamma_{_\mu}\Big)\omega_+\;,
\nonumber\\
&&{\cal O}_{_{\rm se}}\sim {i\over(4\pi)^2}g_{_s}T^am_{_b}\Big(2p_{_\mu}
+/\!\!\!k\gamma_{_\mu}\Big)\omega_+
\label{adeq3}
\end{eqnarray}
in the momentum space. Here, $p,\;k$ are the incoming momenta of the 
external quark $b$ and gauge boson ($\gamma$ or $g$) respectively.
For the effective $\bar{s}b\gamma\;(g)$ vertices $A_{_\mu}^{\gamma\;(g)}(p,k)$,
the corresponding WTIs required by
the $SU(3)_{_c}\times U(1)_{_{em}}$ gauge invariance are written
as
\begin{eqnarray}
&&iee_{_d}\Big(\Sigma(p+k)-\Sigma(p)\Big)=ik\cdot A^\gamma(p,k)\;,
\nonumber\\
&&ig_{_s}T^a\Big(\Sigma(p+k)-\Sigma(p)\Big)=ik\cdot A^g(p,k)\;,
\label{adeq4}
\end{eqnarray}
where $i\Sigma(p)$ represents the sum of amplitudes for the self energy diagrams
(FIG. \ref{fig1}). Expanding the self energy amplitudes in
powers of external momentum to the third order, we have
\begin{eqnarray}
&&i\Sigma(p)={i\over(4\pi)^2}B_{_0}m_{_b}p^2\;,
\label{adeq5}
\end{eqnarray}
where the function $B_{_0}$ only depends on the heavy freedoms which are integrated 
out. In the effective theory, there are two deductions from the WTIs:
\begin{itemize}
\item the effective $\bar{s}b\gamma\;(g)$ vertices can be formulated as
\begin{eqnarray}
&&iA_{_\mu}^\gamma(p,k)={i\over(4\pi)^2}ee_{_d}m_{_b}\Big\{B_{_1}^\gamma\Big(
2p_{_\mu}+/\!\!\!k\gamma_{_\mu}\Big)+B_{_2}^\gamma[/\!\!\!k,\gamma_{_\mu}]\Big\}
\omega_+\;,
\nonumber\\
&&iA_{_\mu}^g(p,k)={i\over(4\pi)^2}g_{_s}T^am_{_b}\Big\{B_{_1}^g\Big(
2p_{_\mu}+/\!\!\!k\gamma_{_\mu}\Big)+B_{_2}^g[/\!\!\!k,\gamma_{_\mu}]\Big\}
\omega_+
\label{adeq6}
\end{eqnarray}
after we expand $A_{_\mu}^{\gamma\;(g)}(p,k)$ in powers of the external
momenta to the second order,
where $B_{_{1,2}}^{\gamma,g}$ are the functions of heavy freedoms only;
\item additionally, $B_{_0}=B_{_1}^\gamma=B_{_1}^g$.
\end{itemize}
The above deductions
can be taken as the criterion to test our calculations.
In Eq. \ref{adeq6}, the functions $B_{_2}^{\gamma,g}$ are proportional 
to the Wilson coefficients of the magnetic and chromo-magnetic dipole 
operators respectively. 

As an application, we will investigate the $CP$ asymmetry of the rare 
decay $B\rightarrow X_{_s}\gamma$ within the framework of MSSM. 

\section{Direct $CP$ violation in $B\rightarrow X_{_s}\gamma$\label{cp}}
\indent\indent
\begin{figure}
\setlength{\unitlength}{1mm}
\begin{center}
\begin{picture}(0,80)(0,0)
\put(-50,-30){\includegraphics{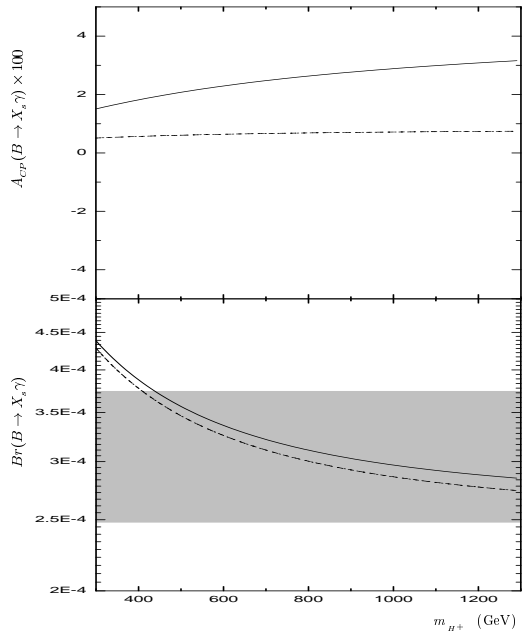}}
\end{picture}
\caption[]{The $CP$ asymmetry and branching ratio of the
inclusive $B\rightarrow X_{_s}\gamma$ decay versus the charged Higgs
mass $m_{_{H^+}}$. Dash-line: theoretical prediction at the one-loop
order, and solid-line: theoretical prediction at the two-loop
order, when $\tan\beta=30,\;|m_{_2}|=300\;{\rm GeV},
\;m_{_{s_R}}=500\;{\rm GeV}$, the other parameters are taken
as in the text. The gray band is the experimental allowed
region for the branching ratio $BR(B\rightarrow X_{_s}\gamma)$ 
at $1\sigma$ deviation.} \label{fig2}
\end{center}
\end{figure}
\begin{figure}
\setlength{\unitlength}{1mm}
\begin{center}
\begin{picture}(0,80)(0,0)
\put(-50,-30){\includegraphics{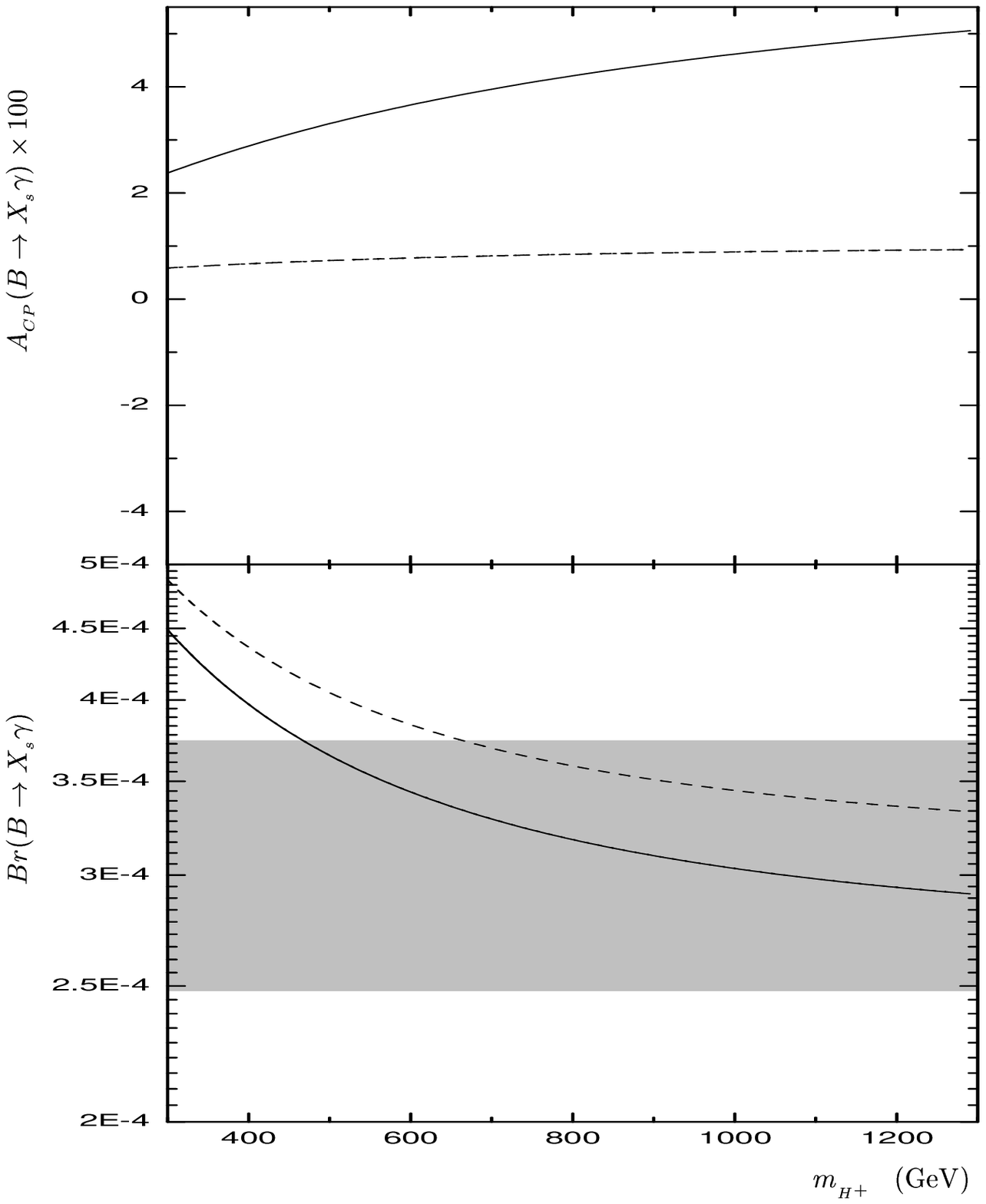}}
\end{picture}
\caption[]{The $CP$ asymmetry and branching ratio of the
inclusive $B\rightarrow X_{_s}\gamma$ decay versus the charged Higgs
mass $m_{_{H^+}}$. Dash-line: theoretical prediction at the one-loop
order, and solid-line: theoretical prediction at the two-loop
order, when $\tan\beta=60,\;|m_{_2}|=300\;{\rm GeV},
\;m_{_{s_R}}=500\;{\rm GeV}$, the other parameters are taken
as in the text. The gray band is the experimental allowed
region for the branching ratio $BR(B\rightarrow X_{_s}\gamma)$ 
at $1\sigma$ deviation.} \label{fig3}
\end{center}
\end{figure}

In the SM, the $CP$ asymmetry of the $B\rightarrow X_{_s}\gamma$ process
\begin{eqnarray}
&&A_{_{CP}}(B\rightarrow X_{_s}\gamma)={\Gamma(\bar{B}\rightarrow
X_{_{\bar s}}\gamma)-\Gamma(B\rightarrow X_{_s}\gamma)\over
\Gamma(\bar{B}\rightarrow X_{_{\bar s}}\gamma)
+\Gamma(B\rightarrow X_{_s}\gamma)}
\label{eq9}
\end{eqnarray}
is calculated to be rather small: $A_{_{CP}}\sim 0.5\%$ \cite{Kagan}.
For experimental data, the recent measurement \cite{Coan} of the $CP$
asymmetry implies the $95\%$ range of
\begin{eqnarray}
&&-0.30\le A_{_{CP}}(B\rightarrow X_{_s}\gamma)\le 0.14\;.
\label{eq10}
\end{eqnarray}
In other words, studies of the direct $CP$ asymmetry in
$B\rightarrow X_{_s}\gamma$ may uncover new sources of
the $CP$ violation which lie outside the SM. Up to the NLO,
the complete theoretical prediction has been presented
in Ref.\cite{Chetyrkin}. In order to eliminate the strong dependence
on the b-quark mass, the branching ratios is usually normalized by
the decay rate of the $B$ meson semileptonic decay:
\begin{eqnarray}
&&{\Gamma(B\rightarrow X_{_s}\gamma)\over\Gamma(B\rightarrow
X_{_c}e\bar{\nu})}={6\alpha\over\pi f(z)}\Big|{V_{_{ts}}^*V_{_{tb}}
\over V_{_{cb}}}C_{_7}(\mu_{_b})\Big|^2\;,
\label{eq11}
\end{eqnarray}
where $f(z)=1-8z+8z^3-z^4-12z^2\ln z$ is the phase-space factor
with $z=(m_{_c}/m_{_b})^2$, and $\alpha=e^2/(4\pi)$ is the electroweak 
fine-structure constant. The $CP$ asymmetry in the rare decay
$B\rightarrow X_{_s}\gamma$ is correspondingly formulated as \cite{Kagan}
\begin{widetext}
\begin{eqnarray}
&&A_{_{CP}}(B\rightarrow X_{_s}\gamma)={\alpha_{_s}(\mu_{_b})\over
|C_{_7}(\mu_{_b})|^2}\Big\{{40\over81}{\bf Im}[C_{_2}(\mu_{_b})
C_{_7}^*(\mu_{_b})]-{4\over9}{\bf Im}[C_{_8}(\mu_{_b})C_{_7}^*(\mu_{_b})]
\nonumber\\&&\hspace{3.2cm}
-{8z\over9}g(z){\bf Im}\Big[\Big(1+{V_{_{us}}^*V_{_{ub}}\over
V_{_{ts}}^*V_{_{tb}}}\Big)C_{_2}(\mu_{_b})C_{_7}^*(\mu_{_b})\Big]\Big\}
\;,
\label{eq12}
\end{eqnarray}
\end{widetext}
with $g(z)=\Big(5+\ln z+\ln^2z-{\pi^2/3}\Big)+\Big(\ln^2z-{\pi^2/3}
\Big)z+\Big(28/9-4/3\ln z\Big)z^2+{\cal O}(z^3)$. The $C_{_2}(\mu_{_b})$
is the Wilson coefficient of the operator ${\cal O}_{_2}=\bar{s}_{_L}
\gamma_\mu q_{_L}\bar{q}_{_L}\gamma^\mu b_{_L}\;(q=c,\;u)$
at the hadronic scale.
From now on we shall assume the value $BR(B\rightarrow
X_{_c}e\bar{\nu})=10.5\%$ for the semileptonic branching ratio,
$\alpha_{_s}(m_{_{\rm z}})=0.118$, $\alpha(m_{_{\rm z}})=1/127$. For the standard
particle masses, we take $m_{_t}=174\;{\rm GeV},\;m_{_b}=4.2\;{\rm GeV}
,\;m_{_{\rm w}}=80.42\;{\rm GeV},\;m_{_{\rm z}}=91.19\;{\rm GeV}$
and $z=m_{_c}/m_{_b}=0.29$. In the CKM matrix, we apply the
Wolfenstein parameterization and set $A=0.85,\;\lambda=0.22,\;
\rho=0.22,\;\eta=0.35$ \cite{Data}.
Without loss of generality, we always assume
the supersymmetric parameters $\mu=A_{_t}e^{-i\pi/2}=100\;{\rm GeV},
\;m_{_3}e^{-i\pi/4}=300\;\;{\rm GeV},\;A_{_s}e^{-i\pi/2}
=m_{_{t_R}}=200\;{\rm GeV}$, $m_{_{t_L}}=m_{_{s_L}}=5\;{\rm TeV}$ 
here. In order to suppress the one-loop EDMs, 
we choose the $\mu$ parameter $CP$ phase $\theta_{_\mu}=0$. As for the
$CP$ phase which is contained in the $SU(2)$ gaugino mass parameter $m_{_2}$,
it is set as $\theta_{_2}=arg(m_{_2})=\pi/4$.

\begin{figure}
\setlength{\unitlength}{1mm}
\begin{center}
\begin{picture}(0,80)(0,0)
\put(-50,-30){\includegraphics{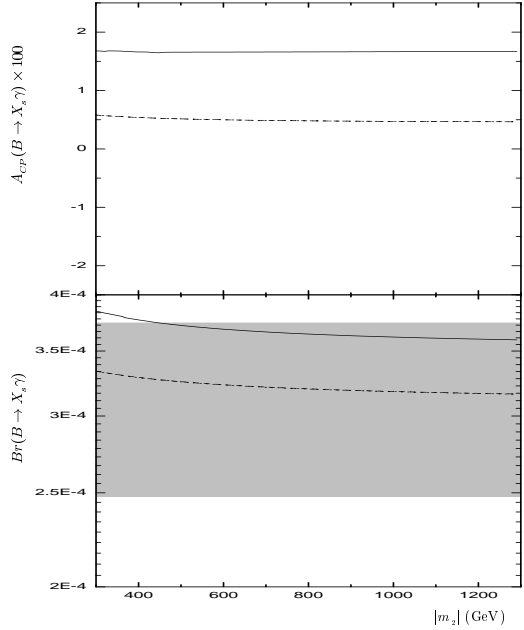}}
\end{picture}
\caption[]{The $CP$ asymmetry and branching ratio of the
inclusive $B\rightarrow X_{_s}\gamma$ decay versus the parameter
$|m_{_2}|$. Dash-line: theoretical prediction at the one-loop
order, and solid-line: theoretical prediction at the two-loop
order, when $\tan\beta=30,\;m_{_{H^+}}=600\;{\rm GeV},
\;m_{_{s_R}}=200\;{\rm GeV}$, the other parameters are taken
as in the text. The gray band is the experimental allowed
region for the branching ratio $BR(B\rightarrow X_{_s}\gamma)$ 
at $1\sigma$ deviation.} \label{fig4}
\end{center}
\end{figure}
\begin{figure}
\setlength{\unitlength}{1mm}
\begin{center}
\begin{picture}(0,80)(0,0)
\put(-50,-30){\includegraphics{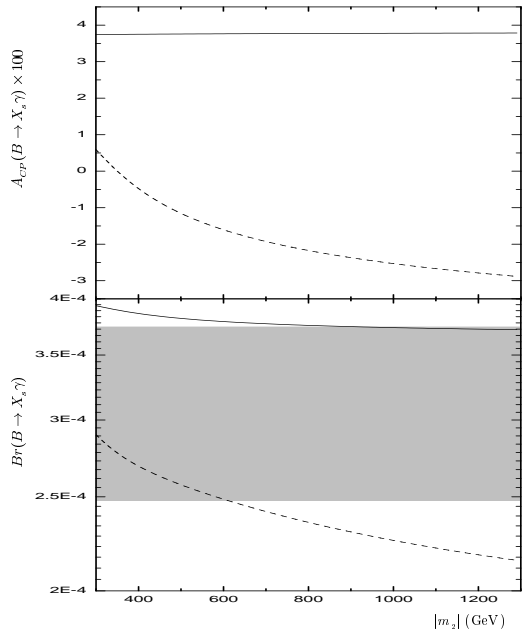}}
\end{picture}
\caption[]{The $CP$ asymmetry and branching ratio of the
inclusive $B\rightarrow X_{_s}\gamma$ decay versus the parameter
$|m_{_2}|$. Dash-line: theoretical prediction at the one-loop
order, and solid-line: theoretical prediction at the two-loop
order, when $\tan\beta=60,\;m_{_{H^+}}=600\;{\rm GeV},
\;m_{_{s_R}}=200\;{\rm GeV}$, the other parameters are taken
as in the text. The gray band is the experimental allowed
region for the branching ratio $BR(B\rightarrow X_{_s}\gamma)$ 
at $1\sigma$ deviation.} \label{fig5}
\end{center}
\end{figure}

Taking $\tan\beta=30,\;|m_{_2}|=300\;{\rm GeV},\;m_{_{s_R}}=500\;{\rm GeV}$,
we plot the $CP$ asymmetry and branching ratio of the inclusive 
$B\rightarrow X_{_s}\gamma$ decay versus the charged Higgs mass
in FIG. \ref{fig2}. Considering the experimental constraint on the 
branching ratio $BR(B\rightarrow X_{_s}\gamma)$ at $1\sigma$ tolerance,
the $CP$ asymmetry including the two-loop corrections can be larger than 
$3\%$, and the one-loop result is smaller than $1\%$ with our chosen
parameters. The choice of parameter space of FIG. \ref{fig3} is identical
with that of FIG. \ref{fig2} except for $\tan\beta=60$. After including
the two-loop corrections, we find that the $CP$ asymmetry can reach $5\%$ with an
increasing of the charged Higgs mass when $\tan\beta=60$, while at the
same time keeping the
branching ratio $BR(B\rightarrow X_{_s}\gamma)$ is within the
$1\sigma$ deviation experimental bound. Since the two-loop correction
is proportional to $\tan\beta$, we can understand
why the differences between the one- and two-loop predictions
of FIG. \ref{fig3} ($\tan\beta=60$) are larger than that of 
FIG. \ref{fig2} ($\tan\beta=30$). From the numerical analysis,
we find that the two-loop corrections to the Wilson coefficients
are still rather smaller than the one-loop results at $\tan\beta=30$,
while the two-loop corrections are comparable
with the one-loop results at $\tan\beta=60$. 
Since the bottom quark Yukawa coupling approximates to 
$1$ at $\tan\beta=60$,
it should be argued whether or not we can safely apply the perturbative
expansion to give the theoretical predictions of physics observables 
for such high $\tan\beta$.

Now, let us study the variance of two-loop results with
the soft $SU(2)$ gaugino mass parameter $|m_{_2}|$. Taking 
$\tan\beta=30,\;m_{_{s_R}}=200\;{\rm GeV},\;
m_{_{H^+}}=600\;{\rm GeV}$, we plot the theoretical predictions for the
$CP$ asymmetry and branching ratio of the inclusive $B\rightarrow X_{_s}\gamma$ 
decay versus the parameter $|m_{_2}|$ in FIG. \ref{fig4} at one- and two-loop order 
respectively. If the theoretical prediction for the branching ratio satisfies with
the experimental bound at $1\sigma$ deviation $$2.48\times10^{-4}\le 
BR(B\rightarrow X_{_s}\gamma)\le3.74\times10^{-4}\;,$$ the $CP$ asymmetry
including the two-loop corrections is about $\sim1.5\%$. The choice of the 
parameter space in FIG. \ref{fig5} is identical with that of FIG. \ref{fig4} 
except for $\tan\beta=60$. In this scenario, the two-loop prediction
on the asymmetry is about $\sim4\%$. Note that the dependence of the two-loop corrections
on the parameter $|m_{_2}|$ is milder than that on the charged Higgs mass
$m_{_{H^+}}$. This fact can be understood as follows: the amplitudes of 
the corresponding triangle diagrams depend on the charged Higgs mass in
form $1/(Q^2-m_{_{H^+}}^2)$ ($Q$ denotes loop momenta $q_1,\;q_2$, or 
${\rm or}\;q_2-q_1$), and depend on the parameter $|m_{_2}|$ through 
the chargino propagator $(/\!\!\!Q-m_{_\chi})/(Q^2-m_{_\chi}^2)$
($m_{_\chi}$ denotes the chargino mass) before the loop momentum 
integration.

In our analysis, we do not compare the exact two-loop analysis with
the HME result since the discussion
has already been presented in Ref. \cite{Borzumati}. 

\section{Conclusions\label{con}}
\indent\indent
In this work, we present the complete two-loop gluino corrections to
inclusive $B\rightarrow X_{_s}\gamma$ decay in explicit $CP$
violating MSSM within large $\tan\beta$ scenarios. Beside the diagrams 
where quark flavor change is mediated by the charged Higgs, we
also include those diagrams in which quark flavor change is mediated by the
charginos. Using loop momentum translation invariant, we formulate
our expressions fulfilling the $SU(3)_{_c}\times U(1)_{_{em}}$
WTIs. From the numerical analysis, we show that the two-loop
corrections to the branching ratio are comparable with the one-loop 
predictions at large $\tan\beta$. Correspondingly, the $CP$ asymmetry
can also reach about $5\%$, which is much larger than that predicted
by the SM.

\begin{acknowledgments}

The work has been supported by the Academy of Finland under the contract 
no.\ 104915.
\end{acknowledgments}
\vspace{2.0cm}
\appendix
\section{Identities among the two-loop scalar integrals\label{appa}}
\indent\indent
Here, we report the identities that are used in the process
of obtaining Eq. \ref{eq7} and Eq.\ref{eq8}, they can be derived from 
the loop momentum translation invariant of the amplitudes. They are
\begin{widetext}
\begin{eqnarray}
&&\int{d^Dq_1\over(2\pi)^D}{d^Dq_2\over(2\pi)^D}{1\over{\cal D}_{_0}
}\Big\{{q_1^2q_1\cdot(q_2-q_1)\over (q_2-q_1)^2-m_{_0}^2}
+{q_1^2q_1\cdot q_2\over q_2^2-m_{_\alpha}^2}\Big\}\equiv0\;,\nonumber\\
\nonumber\\&&
\int{d^Dq_1\over(2\pi)^D}{d^Dq_2\over(2\pi)^D}{1\over{\cal D}_{_0}
}\Big\{{2\over (q_2-q_1)^2-m_{_0}^2}
{(q_1\cdot q_2)^2-q_1^2q_2^2\over D(D-1)}+{q_1\cdot q_2\over D}\Big\}
\equiv0\;,
\nonumber\\&&
\int{d^Dq_1\over(2\pi)^D}{d^Dq_2\over(2\pi)^D}{1\over{\cal D}_{_0}
}\Big\{-{q_1\cdot(q_2-q_1)\over D}
+{2\over q_2^2-m_{_\alpha}^2}{(q_1\cdot q_2)^2-q_1^2q_2^2\over D(D-1)}\Big\}
\equiv0\;,
\nonumber\\&&
\int{d^Dq_1\over(2\pi)^D}{d^Dq_2\over(2\pi)^D}{1\over{\cal D}_{_0}
}\Big\{{1\over (q_2-q_1)^2-m_{_0}^2}
\Big[{D(q_1\cdot q_2)^2-q_1^2q_2^2\over D(D-1)}-{q_1^2q_1\cdot q_2\over D}\Big]
\nonumber\\&&
+{1\over q_2^2-m_{_\alpha}^2}{D(q_1\cdot q_2)^2-q_1^2q_2^2\over D(D-1)}\Big\}
\equiv0\;,
\nonumber\\&&
\int{d^Dq_1\over(2\pi)^D}{d^Dq_2\over(2\pi)^D}{1\over{\cal D}_{_0}
}\Big\{-q_1^2+{2\over D}{q_1^2q_2\cdot(q_2-q_1)\over (q_2-q_1)^2-m_{_0}^2}
+{2\over D}{q_1^2q_2^2\over q_2^2-m_{_\alpha}^2}\Big\}
\equiv0\;,
\nonumber\\&&
\int{d^Dq_1\over(2\pi)^D}{d^Dq_2\over(2\pi)^D}{1\over{\cal D}_{_0}
}\Big\{-q_1\cdot q_2+{q_2^2q_1\cdot(q_2-q_1)\over (q_2-q_1)^2-m_{_0}^2}
+{q_2^2q_1\cdot q_1\over q_2^2-m_{_\alpha}^2}\Big\}
\equiv0\;,
\nonumber\\&&
\int{d^Dq_1\over(2\pi)^D}{d^Dq_2\over(2\pi)^D}{1\over{\cal D}_{_0}
}\Big\{-{2+D\over2}q_2^2+{q_2^2q_2\cdot(q_2-q_1)\over (q_2-q_1)^2-m_{_0}^2}
+{q_2^4\over q_2^2-m_{_\alpha}^2}\Big\}
\equiv0\;,
\nonumber\\&&
\int{d^Dq_1\over(2\pi)^D}{d^Dq_2\over(2\pi)^D}{1\over{\cal D}_{_0}
}\Big\{-q_1\cdot q_2
+{2\over (q_2-q_1)^2-m_{_0}^2}{q_1\cdot q_2(q_2-q_1)^2\over D}
\nonumber\\&&
+{2\over q_2^2-m_{_\alpha}^2}\Big[{q_1\cdot q_2q_2^2\over D}
-{D(q_1\cdot q_2)^2-q_1^2q_2^2\over D(D-1)}\Big]\Big\}\equiv0\;,
\nonumber\\&&
\int{d^Dq_1\over(2\pi)^D}{d^Dq_2\over(2\pi)^D}{1\over{\cal D}_{_0}
}\Big\{{1\over(q_2-q_1)^2-m_{_0}^2}\Big[q_1\cdot q_2q_2\cdot(q_2-q_1)
-{D+1\over2}q_2^2q_1\cdot(q_2-q_1)\Big]
\nonumber\\&&
-{D-1\over2}{q_2^2q_1\cdot q_2\over q_2^2-m_{_\alpha}^2}\Big\}\equiv0\;,
\nonumber\\
&&\int{d^Dq_1\over(2\pi)^D}{d^Dq_2\over(2\pi)^D}{1\over{\cal D}_{_0}
}\Big\{{2\over D}{q_2\cdot(q_2-q_1)\over (q_2-q_1)^2-m_{_0}^2}
+{2\over D}{q_2^2\over q_2^2-m_{_\alpha}^2}-1\Big\}\equiv0\;,
\nonumber\\
&&\int{d^Dq_1\over(2\pi)^D}{d^Dq_2\over(2\pi)^D}{1\over{\cal D}_{_0}}
\Big\{{2\over D}{(q_2-q_1)^2\over (q_2-q_1)^2-m_{_0}^2}
+{2\over D}{q_2\cdot(q_2-q_1)\over q_2^2-m_{_\alpha}^2}-1\Big\}\equiv0\;,
\nonumber\\&&
\int{d^Dq_1\over(2\pi)^D}{d^Dq_2\over(2\pi)^D}{1\over{\cal D}_{_0}
}\Big\{{q_1\cdot(q_2-q_1)\over D}
-{2\over q_2^2-m_{_\alpha}^2}{(q_1\cdot q_2)^2-q_1^2q_2^2\over D(D-1)}\Big\}\equiv0\;,
\nonumber\\&&
\int{d^Dq_1\over(2\pi)^D}{d^Dq_2\over(2\pi)^D}{1\over{\cal D}_{_0}
}\Big\{{q_2\cdot(q_2-q_1)\over D}
+{2\over q_1^2-m_{_a}^2}{(q_1\cdot q_2)^2-q_1^2q_2^2\over D(D-1)}\Big\}\equiv0\;,
\nonumber\\
&&\int{d^Dq_1\over(2\pi)^D}{d^Dq_2\over(2\pi)^D}{1\over{\cal D}_{_0}
}\Big\{-{2+D\over D}q_1\cdot q_2
+{2\over q_1^2-m_{_a}^2}{q_1^2q_1\cdot q_2\over D}
\nonumber\\&&
+{2\over q_2^2-m_{_\alpha}^2}
{D(q_1\cdot q_2)^2-q_1^2q_2^2\over D(D-1)}\Big\}\equiv0\;,
\nonumber\\
&&\int{d^Dq_1\over(2\pi)^D}{d^Dq_2\over(2\pi)^D}{1\over{\cal D}_{_0}
}\Big\{-{2+D\over D}q_1\cdot q_2
+{2\over q_1^2-m_{_a}^2}{D(q_1\cdot q_2)^2-q_1^2q_2^2\over D(D-1)}
\nonumber\\&&
+{2\over q_2^2-m_{_\alpha}^2}{q_1\cdot q_2q_2^2\over D}\Big\}\equiv0\;,
\nonumber\\&&
\int{d^Dq_1\over(2\pi)^D}{d^Dq_2\over(2\pi)^D}{1\over{\cal D}_{_0}
}\Big\{-{2+D\over 2}q_1^2
+{q_1^4\over q_1^2-m_{_a}^2}
+{q_1^2q_1\cdot q_2\over q_2^2-m_{_\alpha}^2}\Big\}\equiv0\;,
\nonumber\\&&
\int{d^Dq_1\over(2\pi)^D}{d^Dq_2\over(2\pi)^D}{1\over{\cal D}_{_0}
}\Big\{-{2+D\over 2}q_2^2
+{q_1\cdot q_2q_2^2\over q_1^2-m_{_a}^2}
+{q_2^4\over q_2^2-m_{_\alpha}^2}\Big\}\equiv0\;,
\nonumber\\
&&\int{d^Dq_1\over(2\pi)^D}{d^Dq_2\over(2\pi)^D}{1\over{\cal D}_{_0}
}\Big\{-q_2\cdot(q_2-q_1)
+{q_1\cdot(q_2-q_1)q_2^2\over q_1^2-m_{_a}^2}
+{q_2\cdot(q_2-q_1)q_2^2\over q_2^2-m_{_\alpha}^2}\Big\}\equiv0\;,
\nonumber\\
&&\int{d^Dq_1\over(2\pi)^D}{d^Dq_2\over(2\pi)^D}{1\over{\cal D}_{_0}
}\Big\{{2\over D}{q_1^2\over q_1^2-m_{_a}^2}
+{2\over D}{q_1\cdot q_2\over q_2^2-m_{_\alpha}^2}-1\Big\}\equiv0\;,
\nonumber\\&&
\int{d^Dq_1\over(2\pi)^D}{d^Dq_2\over(2\pi)^D}{1\over{\cal D}_{_0}}
\Big\{{2\over D}{q_1\cdot q_2\over q_1^2-m_{_a}^2}
+{2\over D}{q_2^2\over q_2^2-m_{_\alpha}^2}-1\Big\}\equiv0\;,
\label{eq13}
\end{eqnarray}
with ${\cal D}_{_0}=((q_2-q_1)^2-m_{_0}^2)(q_1^2-m_{_a}^2)(q_2^2-m_{_\alpha}^2)$,
and $D$ is the time-space dimension. In addition the two-loop vacuum integral
\begin{eqnarray}
&&\int{d^Dq_1\over(2\pi)^D}{d^Dq_2\over(2\pi)^D}{1\over{\cal D}_{_0}},
\label{eq14}
\end{eqnarray}
has been discussed in Ref. \cite{Davydychev}.

\section{Form factors in the two-loop Wilson coefficients\label{appb}}
\indent\indent
\begin{eqnarray}
&&{\cal N}_{_{H(1)}}^\gamma=-{q_1^2q_1\cdot q_2
\over(q_1^2-m_{_{\tilde{s}_i}}^2)^2}-{1\over3}{4(q_1\cdot q_2)^2-q_1^2q_2^2
\over(q_1^2-m_{_{\tilde{s}_i}}^2)(q_2^2-m_{_{H^+}}^2)}
-{q_1\cdot q_2q_2^2\over(q_2^2-m_{_{H^+}}^2)^2}
\nonumber\\
&&\hspace{1.5cm}
+{q_1\cdot q_2\over q_1^2-m_{_{\tilde{s}_i}}^2}+{q_1\cdot q_2\over
q_2^2-m_{_t}^2}\;,\nonumber\\
&&{\cal N}_{_{H(2)}}^\gamma=-{q_1^2\over(q_1^2-m_{_{\tilde{s}_i}}^2)^2}-{q_1\cdot q_2\over
(q_1^2-m_{_{\tilde{s}_i}}^2)(q_2^2-m_{_{H^+}}^2)}
-{q_2^2\over(q_2^2-m_{_{H^+}}^2)^2}+{1\over q_1^2-m_{_{\tilde{s}_i}}^2}
\nonumber\\
&&\hspace{1.5cm}
+{1\over q_2^2-m_{_{H^+}}^2}+{2\over q_2^2-m_{_t}^2}\Big]\;,\nonumber\\
&&{\cal N}_{_{H(1)}}^g=-{q_1^2q_1\cdot q_2
\over(q_1^2-m_{_{\tilde{s}_i}}^2)^2}-{1\over3}{4(q_1\cdot q_2)^2-q_1^2q_2^2
\over(q_1^2-m_{_{\tilde{s}_i}}^2)(q_2^2-m_{_{H^+}}^2)}
-{q_1\cdot q_2q_2^2\over(q_2^2-m_{_{H^+}}^2)^2}
+{17\over8}{q_1\cdot q_2\over q_1^2-m_{_{\tilde{s}_i}}^2}
\nonumber\\
&&\hspace{1.5cm}
-{1\over16}{8q_1\cdot q_2-9q_2^2\over q_2^2-m_{_t}^2}
+{3\over 16}{8q_1\cdot q_2+3q_2^2\over q_2^2-m_{_{H^+}}^2}-{9\over16}
\;,\nonumber\\
&&{\cal N}_{_{H(2)}}^g=-{q_1^2\over(q_1^2-m_{_{\tilde{s}_i}}^2)^2}
-{q_1\cdot q_2\over(q_1^2-m_{_{\tilde{s}_i}}^2)(q_2^2-m_{_{H^+}}^2)}
-{q_2^2\over(q_2^2-m_{_{H^+}}^2)^2}+{1\over q_1^2-m_{_{\tilde{s}_i}}^2}
\nonumber\\
&&\hspace{1.5cm}
+{1\over q_2^2-m_{_{H^+}}^2}-{1\over q_2^2-m_{_t}^2}
-{9\over8}{1\over q_1^2-|m_{_3}|^2}\;,\nonumber\\
&&{\cal N}_{_{\chi_k^\pm(1)}}^\gamma=
{q_1^2q_1\cdot(q_2-q_1)\over(q_1^2-m_{_{\tilde{s}_i}}^2)^2}
-{1\over3}{3q_1^2q_1\cdot q_2-4(q_1\cdot q_2)^2+q_1^2q_2^2\over
(q_1^2-m_{_{\tilde{s}_i}}^2)(q_2^2-m_{_{\chi_k}}^2)}
+{q_1\cdot(q_2-q_1)q_2^2\over(q_2^2-m_{_{\chi_k}}^2)^2}
\nonumber\\
&&\hspace{1.5cm}
-{q_1\cdot(q_2-q_1)\over q_2^2-m_{_{\tilde{t}_j}}^2}
-{2q_1^2-q_1\cdot q_2\over q_2^2-m_{_{\chi_k}}^2}
-{q_1\cdot q_2\over q_2^2-m_{_{\tilde{t}_j}}^2}
\;,\nonumber\\
&&{\cal N}_{_{\chi_k^\pm(2)}}^\gamma=
{q_1^2q_1\cdot q_2\over(q_1^2-m_{_{\tilde{s}_i}}^2)^2}
-{1\over3}{4(q_1\cdot q_2)^2-q_1^2q_2^2\over(q_1^2-m_{_{\tilde{s}_i}}^2)
(q_2^2-m_{_{\chi_k}}^2)}+{q_1\cdot q_2q_2^2\over(q_2^2-m_{_{\chi_k}}^2)^2}
\nonumber\\
&&\hspace{1.5cm}
-{q_1\cdot q_2\over q_1^2-m_{_{\tilde{s}_i}}^2}
-{q_1\cdot q_2\over q_2^2-m_{_{\tilde{t}_j}}^2}
\;,\nonumber\\
&&{\cal N}_{_{\chi_k^\pm(3)}}^\gamma=
{q_1^2q_2\cdot(q_2-q_1)\over(q_1^2-m_{_{\tilde{s}_i}}^2)^2}
+{1\over3}{3q_1\cdot q_2q_2^2-4(q_1\cdot q_2)^2+q_1^2q_2^2\over
(q_1^2-m_{_{\tilde{s}_i}}^2)(q_2^2-m_{_{\chi_k}}^2)}
+{q_2\cdot(q_2-q_1)q_2^2\over(q_2^2-m_{_{\chi_k}}^2)^2}
\nonumber\\
&&\hspace{1.5cm}
-{q_2\cdot(q_2-q_1)\over q_1^2-m_{_{\tilde{s}_i}}^2}
-{q_2^2\over q_2^2-m_{_{\chi_k}}^2}-{2q_2^2-q_1\cdot q_2
\over q_2^2-m_{_{\tilde{t}_j}}^2}-2
\;,\nonumber\\
&&{\cal N}_{_{\chi_k^\pm(4)}}^\gamma=
-{q_1^2\over(q_1^2-m_{_{\tilde{s}_i}}^2)^2}-{q_2^2\over(q_2^2-m_{_{\chi_k}}^2)^2}
-{q_1\cdot q_2\over(q_1^2-m_{_{\tilde{s}_i}}^2)(q_2^2-m_{_{\chi_k}}^2)}
\nonumber\\
&&\hspace{1.5cm}
+{1\over q_1^2-m_{_{\tilde{s}_i}}^2}-{2\over q_2^2-m_{_{\chi_k}}^2}
-{2\over(q_2-q_1)^2-m_{_t}^2}\;,\nonumber\\
&&{\cal N}_{_{\chi_k^\pm(1)}}^g=
{q_1^2q_1\cdot(q_2-q_1)\over(q_1^2-m_{_{\tilde{s}_i}}^2)^2}
-{1\over3}{3q_1^2q_1\cdot q_2-4(q_1\cdot q_2)^2+q_1^2q_2^2\over
(q_1^2-m_{_{\tilde{s}_i}}^2)(q_2^2-m_{_{\chi_k}}^2)}
+{q_1\cdot(q_2-q_1)q_2^2\over(q_2^2-m_{_{\chi_k}}^2)^2}
\nonumber\\
&&\hspace{1.5cm}
-{17\over8}{q_1\cdot(q_2-q_1)\over q_1^2-m_{_{\tilde{s}_i}}^2}
-{1\over16}{7q_1\cdot q_2-q_1^2+9q_2^2\over q_2^2-m_{_{\chi_k}}^2}
-{9\over16}{q_2\cdot(q_2-q_1)\over q_2^2-m_{_{\tilde{t}_j}}^2}
\;,\nonumber\\
&&{\cal N}_{_{\chi_k^\pm(2)}}^g=
{q_1^2q_1\cdot q_2\over(q_1^2-m_{_{\tilde{s}_i}}^2)^2}
+{1\over3}{4(q_1\cdot q_2)^2-q_1^2q_2^2\over(q_1^2-m_{_{\tilde{s}_i}}^2)
(q_2^2-m_{_{\chi_k}}^2)}+{q_1\cdot q_2q_2^2\over(q_2^2-m_{_{\chi_k}}^2)^2}
\nonumber\\
&&\hspace{1.5cm}
-{17\over8}{q_1\cdot q_2\over q_1^2-m_{_{\tilde{s}_i}}^2}
-{3\over16}{8q_1\cdot q_2+3q_2^2\over q_2^2-m_{_{\chi_k}}^2}
+{1\over16}{8q_1\cdot q_2-9q_2^2\over q_2^2-m_{_{\tilde{t}_j}}^2}
+{9\over8}\;,\nonumber\\
&&{\cal N}_{_{\chi_k^\pm(3)}}^g=
{q_1^2q_2\cdot(q_2-q_1)\over(q_1^2-m_{_{\tilde{s}_i}}^2)^2}
+{1\over3}{3q_1\cdot q_2q_2^2-4(q_1\cdot q_2)^2+q_1^2q_2^2\over
(q_1^2-m_{_{\tilde{s}_i}}^2)(q_2^2-m_{_{\chi_k}}^2)}
+{q_2\cdot(q_2-q_1)q_2^2\over(q_2^2-m_{_{\chi_k}}^2)^2}
\nonumber\\
&&\hspace{1.5cm}
-{q_2\cdot(q_2-q_1)\over q_1^2-m_{_{\tilde{s}_i}}^2}
+{1\over8}{12q_1\cdot q_2-13q_2^2\over q_2^2-m_{_{\chi_k}}^2}
+{1\over16}{7q_2^2-8q_1\cdot q_2\over q_2^2-m_{_{\tilde{t}_j}}^2}
+{9\over8}{q_2\cdot(q_2-q_1)\over q_1^2-|m_{_3}|^2}
+{1\over8}
\;,\nonumber\\
&&{\cal N}_{_{\chi_k^\pm(4)}}^g=
-{q_1^2\over(q_1^2-m_{_{\tilde{s}_i}}^2)^2}
-{q_1\cdot q_2\over(q_1^2-m_{_{\tilde{s}_i}}^2)(q_2^2-m_{_{\chi_k}}^2)}
+{q_2^2\over(q_2^2-m_{_{\chi_k}}^2)^2}
+{1\over q_1^2-m_{_{\tilde{s}_i}}^2}
\nonumber\\
&&\hspace{1.5cm}
+{1\over q_2^2-m_{_{\chi_k}}^2}-{9\over8}{1\over q_1^2-|m_{_3}|^2}
-{1\over8}{1\over(q_2-q_1)^2-m_{_t}^2}\;.
\label{eq15}
\end{eqnarray}
\end{widetext}

\end{document}